\documentclass[aps,pra,showpacs,twocolumn,groupedaddress]{revtex4}
\usepackage{graphicx,amsmath,amssymb}
\begin{document}
\newtheorem{Theorem}{Theorem}

\title{A modified quantum adiabatic evolution for the Deutsch-Jozsa problem}

\author{Zhaohui Wei and Mingsheng Ying}

\affiliation{ State Key Laboratory of Intelligent Technology and
Systems, Department of Computer Science and Technology, Tsinghua
University, Beijing, China, 100084}

\begin{abstract}

Deutsch-Jozsa algorithm has been implemented via a quantum adiabatic
evolution by S. Das et al. [Phys. Rev. A 65, 062310 (2002)]. This
adiabatic algorithm gives rise to a quadratic speed up over
classical algorithms. We show that a modified version of the
adiabatic evolution in that paper can improve the performance to
constant time.

\end{abstract}
\pacs{03.67.Lx}

\maketitle

Quantum computation has attracted a great deal of attention in the
recent years, because some quantum algorithms show that the
principles of quantum mechanics can be used to greatly enhance the
efficiency of computation. Quantum algorithms that have been
invented include Deutsch-Jozsa algorithm \cite{CEMM98,DJ92}, Shor's
algorithm \cite{SHOR94} and Grover's algorithm \cite{GROVER97}. The
first one is what we will discuss in this paper.

Assume that we have a boolean function of the form
$$f:\{0,1\}^{n}\rightarrow\{0,1\},$$
and it has been known that the function is either constant (i.e.,
all outputs are identical) or balanced (i.e., has an equal number of
0's and 1's as outputs). Our task is to decide whether it is
constant or not. To solve this problem, any deterministic classical
algorithm needs $2^{n}/2+1$ evaluation of the function $f$, while
Deutsch-Jozsa algorithm \cite{CEMM98,DJ92} shows that a quantum
computer can solve the same problem with one evaluation.

When these algorithms mentioned above were invented, they were
implemented using quantum circuits involving a sequence of unitary
operators. We say these algorithms stay within the standard paradigm
of quantum computation. Recently, another novel paradigm based on
quantum adiabatic evolution has been proposed for quantum
computation \cite{FGGS00}. In a quantum adiabatic algorithm, the
state of the quantum register evolves under a hamiltonian that
varies continuously and slowly. At the beginning, we let the state
of the system in the ground state of the initial hamiltonian. If the
hamiltonian of the system evolves slowly enough, the quantum
adiabatic theorem guarantees that the final state of the system will
differ from the ground state of the final hamiltonian by a
negligible amount. If we encode the solution of the algorithm in the
ground state of the final hamiltonian, after the quantum adiabatic
evolution we can get the solution with high probability by measuring
the final state.

Some quantum algorithms have been reproduced by adiabatic
evolutions. In \cite{DKK02} S. Das et al. solved the Deutsch-Jozsa
problem by an adiabatic evolution. This adiabatic evolution gives
rise to a quadratic speed up over classical algorithms. Recently,
the experimental implementation of this adiabatic algorithm has been
reported by A. Mitra et al \cite{MGDPK05}. However, compared to the
standard quantum algorithm for this problem, this adiabatic
algorithm is not so good and needs to be improved. In this paper, we
propose a modified adiabatic evolution for the Deutsch-Jozsa problem
based on the one in \cite{DKK02}, and we will show that the running
time of the modified adiabatic evolution is constant. (We have noted
that in \cite{SL05} M. S. Sarandy and D. A. Lidar have obtained the
same result via a different adiabatic evolution.)

For convenience of the readers, we present an overview of adiabatic
algorithms. Suppose $H_0$ and $H_T$ are the initial and the final
Hamiltonians of the system. Suppose $|\alpha\rangle$, the ground
state of $H_0$, is the initial state of the system and
$|\beta\rangle$, the ground state of $H_T$, is the final state that
encodes the solution. Then we let the system vary under the
following time dependent Hamiltonian:
\begin{equation}
H(t)=(1-s)H_0+sH_T,
\end{equation}
where $s=s(t)$ is a monotonic function with $s(0)=0 $ and $s(T)=1$
($T$ is the running time of the evolution). Let $|E_0,t\rangle$ and
$|E_1,t\rangle$ be the ground state and the first excited state of
the Hamiltonian at time t, and let $E_0(t)$ and $E_1(t)$ be the
corresponding eigenvalues. The adiabatic theorem \cite{LIS55} shows
that we have
\begin{equation}
|\langle E_0,T|\psi(T)\rangle|^{2}\geq1-\varepsilon^2,
\end{equation}
provided that
\begin{equation}
\frac{D_{max}}{g_{min}^2}\leq\varepsilon,\ \ \ \ 0<\varepsilon\ll1,
\end{equation}
where $g_{min}$ is the minimum gap between $E_0(t)$ and $E_1(t)$
\begin{equation} g_{min}=\min_{0\leq t \leq T}[E_1(t)-E_0(t)],
\end{equation}
and $D_{max}$ is a measurement of the evolving rate of the
Hamiltonian
\begin{equation}
D_{max}=\max_{0\leq t \leq
T}|\langle\frac{dH}{dt}\rangle_{1,0}|=\max_{0\leq t \leq T}|\langle
E_1,t|\frac{dH}{dt}|E_0,t\rangle|.
\end{equation}

In the adiabatic evolution of \cite{DKK02}, the initial and the
final Hamiltonians are

\begin{equation}
H_0=I-|\alpha\rangle\langle\alpha|,
\end{equation}
\begin{equation}
H_T=I-|\beta\rangle\langle\beta|,
\end{equation}
where
\begin{equation}
|\alpha\rangle=\frac{1}{\sqrt{N}}\sum\limits_{i=0}^{N-1}{|i\rangle},\
\ N = 2^n,
\end{equation}
\begin{equation}
|\beta\rangle=\mu|0\rangle+\frac{\nu}{\sqrt{N-1}}\sum\limits_{i=1}^{N-1}{|k\rangle},
\end{equation}
with
\begin{equation}
\mu=\frac{1}{N}|\sum\limits_{x\in\{0,1\}^n}^{}{(-1)^{f(x)}}|,
\end{equation}
\begin{equation}
\nu=1-\mu.
\end{equation}
To solve the Deutsch-Jozsa problem, after the adiabatic evolution
ends we measure the final state of the system. If the measurement
yields $|0\rangle$, $f(x)$ is constant and if the measurement
doesn't yield $|0\rangle$, $f(x)$ is balanced.

In \cite{DKK02}, S. Das et al. showed that the running time of the
local adiabatic evolution \cite{RC02} above is
\begin{equation}
T = O(\sqrt{N}).
\end{equation}
So, the performance of the adiabatic algorithm for this problem is
related to $n$. However, using standard quantum computational
techniques to solve the Deutsch-Jozsa problem \cite{CEMM98,DJ92},
the quantum computer needs only one evaluation of the function $f$
no matter how big $n$ is. We may guess that the adiabatic evolution
can be improved. In fact, we can do this by modifying the adiabatic
algorithm as follows.

In the modified adiabatic evolution, we choose the new final state
\begin{equation}
|\beta\rangle=\frac{\mu}{\sqrt{N/2}}\sum\limits_{k=0}^{N/2-1}{|2k\rangle}+\frac{\nu}{\sqrt{N/2}}\sum\limits_{i=0}^{N/2-1}{|2i+1\rangle},
\end{equation}
and $H_0$,$H_T$,$|\alpha\rangle$,$\nu$,$\mu$ ,$H(s)$ don't change.
Furthermore, we assume that $s(t)$ is linear in t, $s=t/T$.
Obviously, if the final state of the system is $|\beta\rangle$, we
can decide whether the function $f(x)$ is balanced or constant by
measuring the system after the adiabatic evolution. If the
measurement yields $|i\rangle$ and i is even, $f(x)$ is constant. If
the measurement yields $|i\rangle$ and i is odd, $f(x)$ is balanced.
Now we prove that the running time of the modified adiabatic
evolution is constant.

Firstly, it's easy to prove that
\begin{equation}
|\langle\alpha|\beta\rangle|=\frac{1}{\sqrt{2}}.
\end{equation}
According to the claim 7 of \cite{AT03} (see also \cite{ZM05}), we
know that
\begin{equation}
g_{min}=\frac{1}{\sqrt{2}}.
\end{equation}
On the other hand,
\begin{equation}
D_{max}=\max_{0\leq t \leq
T}|\langle\frac{dH}{dt}\rangle_{1,0}|\leq\frac{1}{T}\max_{0\leq t
\leq T}\parallel\frac{dH}{ds}\parallel.
\end{equation}
Substituting $\parallel\frac{dH}{ds}\parallel=\parallel
H_T-H_0\parallel\leq2$ into Eq.(16), we get
\begin{equation}
D_{max}\leq\frac{2}{T}.
\end{equation}
Substituting Eqs. (15), and (17) in Eq. (3), we have
\begin{equation}
T\geq \frac{4}{\varepsilon}.
\end{equation}
That is to say, the running time of the modified adiabatic evolution
is independent of $N$.

In conclusion, we have shown that a modified version of the
adiabatic evolution of \cite{DKK02} can solve the Deutsch-Jozsa
problem in constant time. This is consistent with the result of
standard quantum algorithm based on quantum gates.

We would like to thank Ji Zhengfeng, M. S. Sarandy, and D. A. Lidar
for useful discussions.

\end{document}